# Complexity Analysis of Unsaturated Flow in Heterogeneous Media Using a Complex Network Approach

Hamed.O.Ghaffari, Mamdou Fall and Erman.Evgin

Dept. Civil Engineering, University of Ottawa, Ottawa, Ontario, Canada

**Abstract**: In this study, we investigate the complexity of two-phase flow (air/water) in a heterogeneous soil sample by using a complex network theory .Based on the different similarity measurements (i.e., correlation, Euclidean metrics) over the emerged patterns from the evolution of saturation of non-wetting phase of a multi-heterogeneous soil sample, the emerged complex networks are recognized. Understanding of the properties of complex networks (such degree distribution, mean path length, clustering coefficient) can be supposed as a way to analysis of variation of saturation profiles structures (as the solution of finite element method on the coupled PDEs) where complexity is coming from the changeable connection and links between assumed nodes. Also, the path of evolution of the supposed system will be illustrated on the state space of networks either in correlation and Euclidean measurements. The results of analysis showed in a closed system the designed complex networks approach to a small world network where the mean path length and clustering coefficient are low and high, respectively. As another result, the evolution of macro -states of system (such mean velocity of air or pressure) can be scaled with characteristics of structure complexity of saturation. In other part, we tried to find a phase transition criterion based on the variation of non-wetting phase velocity profiles over a network which had been constructed over correlation distance.

## 1. Introduction

Complexity of flow in porous media–especially in multiphase state- is the result of multi dimensional (geological) heterogeneity of particles where each element of system (or group of elements) based on nonlinear mutual relation with gravity, capillarity and pressure gradient shows its behaviour. In analysis of such rich complex behaviour apart of physical simplification to solve the problem, the possible uncertainties in different format (spontaneously or non-spontaneously) will induce other facet of complexity. This type of complexity has allocated considerable literature where uncertainty (i.e. stochastic or fuzzy) in direct form construct stochastic differential equations or fuzzy differential equation while in indirect branch uncertainty acts basically over collected information from system and gives inferences associated with in(i.e. fuzzy) or on(i.e. rough set) boundaries of system [1],[17],[19].

Along this complexity, the interaction of particles which their dynamic directly are governed by forces (agents) exhibits another complexity (absolutely coupling with uncertainty and controlling equations): structure and topology based complexity. Undoubtedly, the later complexity has an effective role in recognizing of state of each particle where the interactions of the (local) states with each other and external forces give practical measurement of the system. In other world, dynamical superposition of complexity in external forces, uncertainty in behavior and interaction of particles may give a general evolution path of a system [1, 4]. The point that must be considered is the role of local interactions of particles (in spatial or /and temporal form) on the evolution of system so that such locality in different computational based methods (i.e., game theory, cellular automata, lattice Boltzmann methods, finite difference solutions, etc) has been taken in to account. However, the idea behind the mentioned methods (as the basic and initial idea) is considering the function of immediate neighborhoods[2-3].

Understanding of structures Complexity in variation and dynamic of a physical system can be evaluated by constructing links (may be with weights) over nodes (as particles or agents). This interpretation of topology complexity during the last years has opened a new perspective to traditional graph theory, called complex networks [2]. Picturing, modeling and evaluation in a simple and intuitive way are some of the discriminated features of complex networks [3, 4]. Based on the mentioned description of the topology structure over space and/or time, possible explicit or implicit of agents /governing equations within or behind of complexity can lead to networks on differential equations (NODs-NPDs) or linguistic rules such fuzzy rules (NFRs) [4] .Complex networks have been developed in the several fields of science and engineering for example social, information, technological, biological and earthquake networks are the main distinguished networks [5-12]. Regard to the geosciences field, during past decade, some considerable efforts have been done on the complex earthquake networks (generally recursive events [7, 10] and climate dynamic networks. The dynamical growth of a thin film surface (and in a nano-structural way) and evolution of a rock joint (complex aperture network-millimeter scale) by this approach have been devised [8, 10].

The aim of this study is to analysis of complexity evolution of saturation (of air) profiles in a closed heterogeneous soil sample (under mildly linear evolution of gas) either in spatial or temporal way. In this way based on two metric spaces, firstly along each sampled time steps, networks will be constructed. Characterization of the appeared networks and comparison of network properties with each other and other distinguished mean properties of system (such pressure and velocity) will be the next step of this study. In fact we are searching a suitable space in where emerged properties of networks over one parameter can scaled with other dimension (attribute) of system. In other word, the question "is there any hidden space so that structure evolution of one property over that space is related to another attribute of same system". Also, we will make state spaces of complex networks where show the path of evolution of system where are dissimilar for evaluated distances. The organization of this paper is as follow: in the second part we will introduce the model parameters and accompanied governing equations; the next section will cover a brief on complex networks properties and the results of the designated networks on saturation and velocity profiles will be completed this section.

## 2. Advection flow in porous media

Generally, the process of displacement of wetting/non-wetting phase is mainly affected by the properties of the permeable medium, and fluids in single-phase both in homogeneous and in heterogeneous media. In two-phase immiscible flows, interactions between the permeable medium and the fluids also affect the fluid flow paths. Because flow dynamics depend on a combination of conditions such as heterogeneity, moisture content, and chemistry, the resulting transient flow and transport are usually complex. A large number of numerical methods have been developed to model two-phase flow in heterogeneous media. The finite difference and finite volume methods are the general frameworks for numerical simulation for the study of fluid flow in very large problems. It has been proved that the finite difference method, however, is strongly influenced by the mesh quality and orientation, which makes the method unattractive for unstructured gridding [13-14].

Finite element methods (in different forms), have been used to model single-phase and two-phase flow [13-17] in fractured media and heterogonous permeable media with different capillarity pressures. In this part of study, based on the transition modeling of immiscible two-phase flow, the role of permeability, porosity and the width of pore-size distribution in a small sample of soil using FEMLAB [18] will be evaluated, while non-wetting phase (air) changes with time, mildly. In all of this study, we assume that the porous medium is a non-deformable (constant porosity) and that cross-product permeability terms associated with the viscous drag tensor can be neglected. The general form of the two-fluid flow equations (without source–sink terms) is described by the two-fluid, volume-averaged momentum and continuity equations [15-17]:

$$\mathbf{q}_w = -\frac{\mathbf{k}_w}{\mu_w}[\nabla P_w + \rho_w \mathbf{g}] \quad (1)$$

$$\phi \frac{\partial(\rho_w S_w)}{\partial t} + \nabla.(\rho_w \mathbf{q}_\mathbf{w}) = 0 \quad (2)$$

$$\mathbf{q}_{nw} = -\frac{\mathbf{k}_{nw}}{\mu_{nw}}.[\nabla p_{nw} + \rho_{nw} \mathbf{g}] \quad (3)$$

$$\phi \frac{\partial(\rho_{nw} S_{nw})}{\partial t} + \nabla.(\rho_{nw} \mathbf{q}_{n\mathbf{w}}) = 0 \quad (4)$$

In Eqs (1-4), the subscripts *w* and *nw* denote the wetting and non-wetting fluids, respectively; $P_i(i=w,nw)$, $S_i$ , $\mathbf{q}_i$, $\mathbf{g}$ , $\mu_i$ and $\mathbf{k}_i$ denote pressure ,degree of saturation relative to the porosity $\phi$ , the flux density vector , the gravitational acceleration vector, dynamic viscosity , density and the effective permeability tensor, respectively. The effective permeability can be defined as the relation between intrinsic permeability ( $\mathbf{k}$ ) and relative permeability ( $k_{ri}$ ): $\mathbf{k}_i = k_{ri}\mathbf{k}$ .

With definition of volumetric fluid content as $\theta_i = \phi S_i$ we can write:

$$S_w + S_{nw} = 1$$
$$\theta_w + \theta_{nw} = \phi \tag{5}$$

Assuming one-dimensional vertical flow and that the wetting fluid is incompressible, substitution of Eq. (1) into Eq. (2) gives:

$$\frac{\partial \theta_w}{\partial t} = \frac{\partial}{\partial z}\left[\frac{k_w}{\mu_w}\left(\frac{\partial P_w}{\partial z} + \rho_w g\right)\right] \tag{6}$$

When replacing fluid pressures and capillary pressure ( $P_c = P_{nw} - P_w$ ) with pressure head $h_i = P_i / \rho_{H_2O} \cdot g$ and defining the hydraulic conductivity of fluid *i* by $K_i = \frac{k_i \rho_{H_2O} \cdot g}{\mu_i}$, transient flow of the wetting fluid by considering the capillary capacity ( $C_w = d\theta_w / dh_c$ ) is described by :

$$C_w\left(\frac{\partial h_{nw}}{\partial t} - \frac{\partial h_w}{\partial t}\right) = \frac{\partial}{\partial z}\left[K_w\left(\frac{\partial h_{nw}}{\partial z} + 1\right)\right] \tag{7}$$

Substitution of Eq.(3) in Eq.(4) yields :

$$\frac{\partial(\rho_{nw}\theta_{nw})}{\partial t} = \frac{\partial}{\partial z}\left[\frac{\rho_{nw} k_{nw}}{\mu_{nw}}\left(\frac{\partial P_{nw}}{\partial z} + \rho_{nw} g\right)\right] \tag{8}$$

For air, the density of non –wetting fluid has a dependency to the pressure head:

$$\rho_{nw} = \rho_{0,nw} + \left(\frac{\rho_{0,nw}}{h_0}\right) h_{nw} \tag{9}$$

where $\rho_{0,nw}$, $\rho_{0,nw}$ are the reference density and pressure head (at atmospheric pressure), respectively. The ratio of $\frac{\rho_{0,nw}}{h_0}$ is defined as the compressibility $\lambda(\lambda = 1.24 \times 10^{-6}\ g / cm^4)$. Then, the flow of non-wetting fluid can be concluded:

$$\{(\phi - \theta_w)\lambda - \rho_{nw} C_w\}\frac{\partial h_{nw}}{\partial t} + \rho_{nw} C_w \frac{\partial h_w}{\partial t} = \frac{\partial}{\partial z}\left[\rho_{nw} K_{nw}\left(\frac{\partial h_{nw}}{\partial z} + \frac{\rho_{nw}}{\rho_{H_2O}}\right]\right. \tag{10}$$

Equations (7) and (10) for the pre-defined boundary conditions can be solved simultaneously (coupled PDES). By functional description of the capillary pressure–saturation, $h_c(S_w)$, and permeability functions, $k_i(S_w)$, the evolution of wetting and non-wetting phases distribution can be estimated.. In this study, we use the van Genuchten equation (VG) [21]:

$$S_{ew} = [1 + (\alpha h_c)^n]^{-m} \tag{11}$$

where $S_{ew}$ denotes the effective saturation of the wetting fluid, $S_{ew} = {(\theta_w - \theta_{wr})}/{(\theta_{ws} - \theta_{wr})}$, where $\theta_{wr}$ and $\theta_{wr}$ are the saturated and residual wetting fluid saturation, respectively; $\alpha$ and $n$ are fitting parameters, that are inversely proportional to the non-wetting fluid entry pressure value and the width of pore-size distribution, respectively. We assume that $m = 1 - {1}/{n}$, and that the effective saturation of the non-wetting fluid ($S_{en}$) is derived from $S_{en} = 1 - S_{ew}$.

The capillary pressure–saturation function can be considered a static soil property, while the permeability function is a hydrodynamic property describing the ability of the soil to conduct a fluid. The basic assumption behind Capillary pressure–permeability prediction models are from conceptual models of flow in capillary tubes combined with pore-size distribution knowledge which are derived from the capillary pressure–saturation relationship. A typical representation of this type of model follows Mualem formulation [22]:

$$k_{rw} = S_{ew}^{\eta} \left[ \frac{\int_0^{S_e} \frac{dS_e}{h_c}}{\int_0^1 \frac{dS_e}{h_c}} \right]^2 \tag{12a}$$

$$k_{rn} = (1 - S_{ew})^{\eta} \left[ \frac{\int_{S_e}^1 \frac{dS_e}{h_c}}{\int_0^1 \frac{dS_e}{h_c}} \right]^2 \tag{12b}$$

Combining the van Genuchten capillary pressure–saturation Eq. (11) with the Mualem (VGM) model-with introducing new parameter the tortuosity parameter ($\eta$) gives permeability functions as follows by [15,22]:

$$k_{rw} = \frac{k_w}{k} = S_{ew}^{\eta}\left[1-(1-S_{ew}^{\frac{1}{m}})^m\right]^2 \quad (13a)$$

$$k_{rnw} = \frac{k_{nw}}{k} = (1-S_{ew})^{\eta}\left[1-S_{ew}^{\frac{1}{m}}\right]^{2m} \quad (13b)$$

Combination Eqs. (8) and (10), the boundary and initial conditions (Fig. 1), and the constitutive relationships in Eq.(12-13) constitute the mathematical model of the assumed system. In the upper bound of the geometry (fig. 1), a slight growth of the air pressure in 0 to 1hour (see Fig. 2) will be increased where the lower bound of the sample is in-transferable to non-wetting phase (Fig. 1). The experiment injects air over the surface of a laboratory column filled with water and sand. The incoming air (the nonwetting phase for this fluid pair) forces the water (the wetting phase) toward the outlet at the base of the column. At the inlet, air pressure increases by steps in time, and no water exits through the column top. In moving to the outlet, the water passes through a disc that is impermeable to air flow. Neither the air nor the water can pass through the vertical column walls. The water pressure at the outlet, which changes in time, corresponds to the height of fluid rise in a receiving buret. The column has a total length of 8.34 cm, a 6-cm radius, and the disk is 0.74 cm thick [15]. Initial non-wetting phase pressure head is assumed 0.2 (m water) and at the water outlet, the fluid level increases linearly in time from 0 m to 0.1 m. The employed parameters for the sample come from Lincoln sand (88.6% sand, 9.4% silt and 2.0% clay). The constant parameters for our model are as follow (see figure 3 for uncorrelated normal distribution of random parameters):

$$\left\{\begin{array}{l} \theta_r(cm^3cm^{-3}) = 0.0210 \\ \theta_s = [\theta_s]; \\ k(cm^2\times 10^{-9}) = [k_{ij}] \\ \alpha(cm^{-1}) = 0.0189 \\ n = [n_{ij}] \end{array}\right\}$$

in which $\theta_s$, $k$ and $n$ will follow an uncorrelated Gaussian distribution drawn in Figure 3.

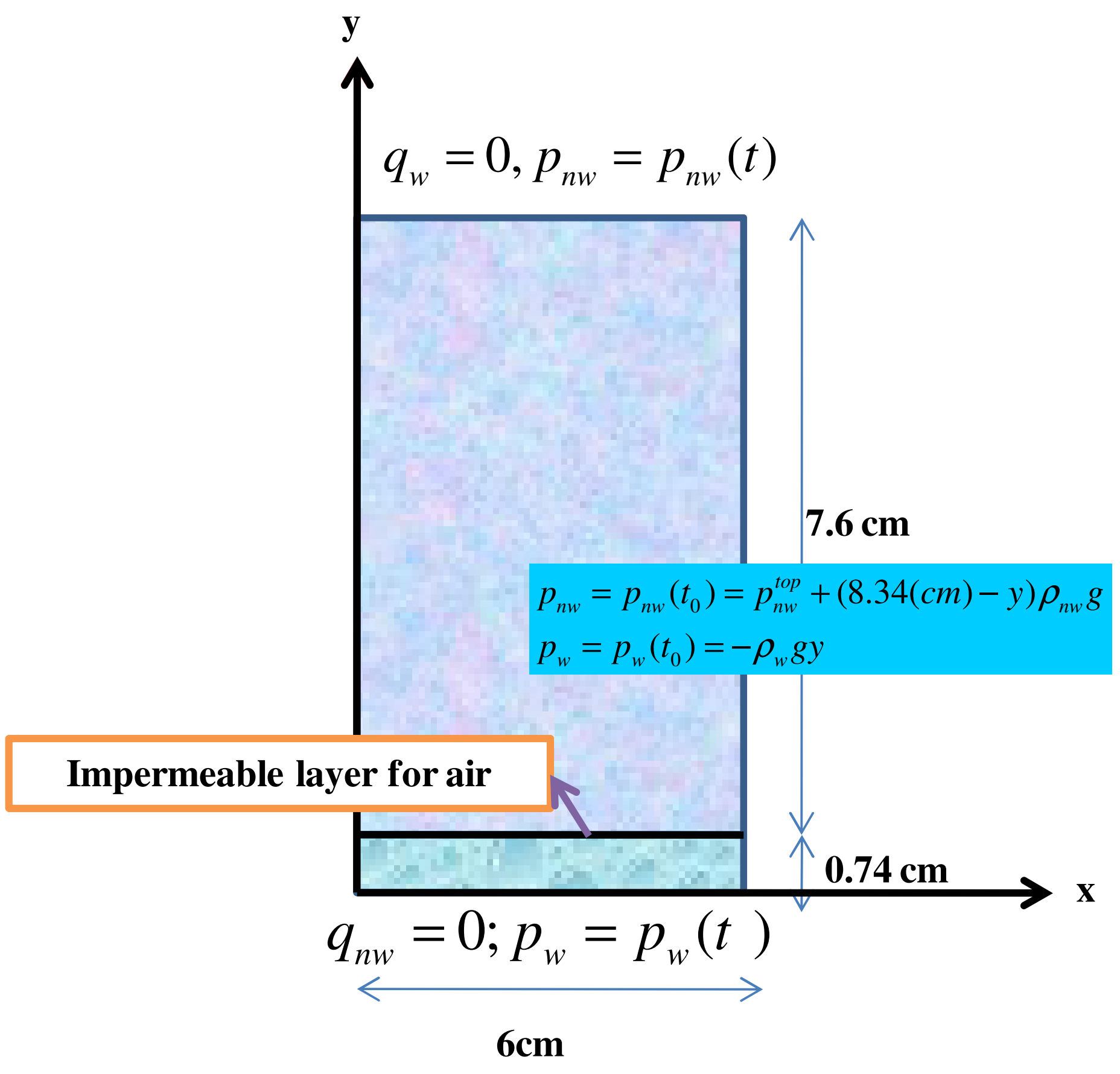

Figure 1. Schematic representation of boundary and initial conditions [15,18].

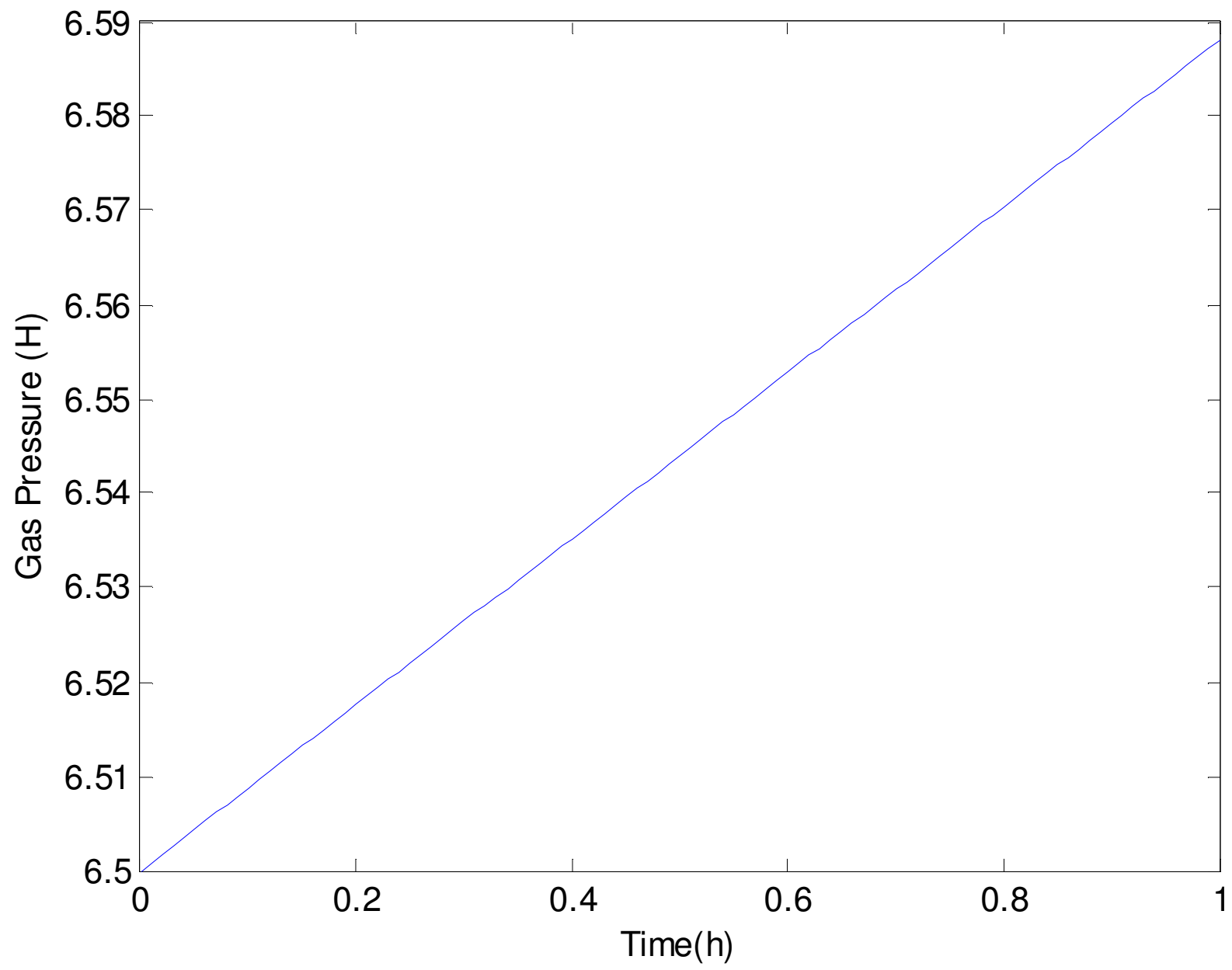

Figure 2. Gas evolution transition on a supposed sample

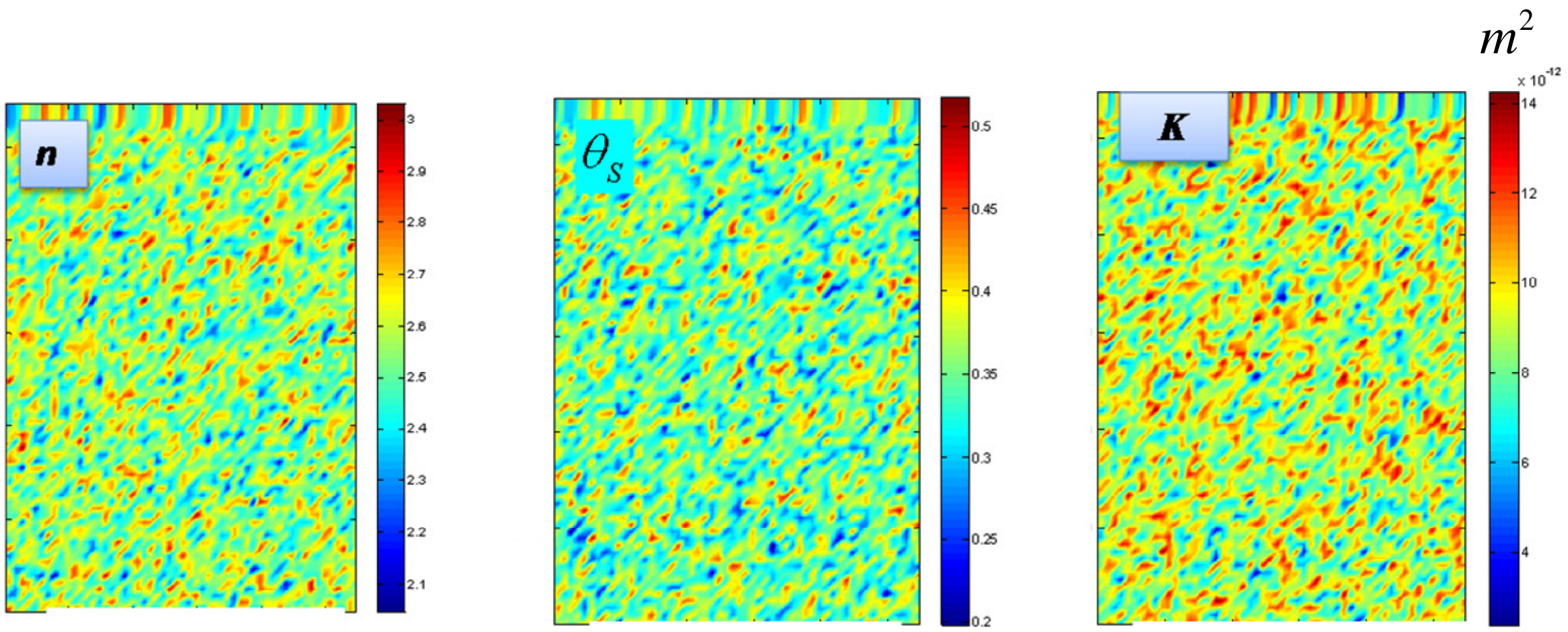

Figure 3. Intrinsic Permeability variation based on a normal distribution

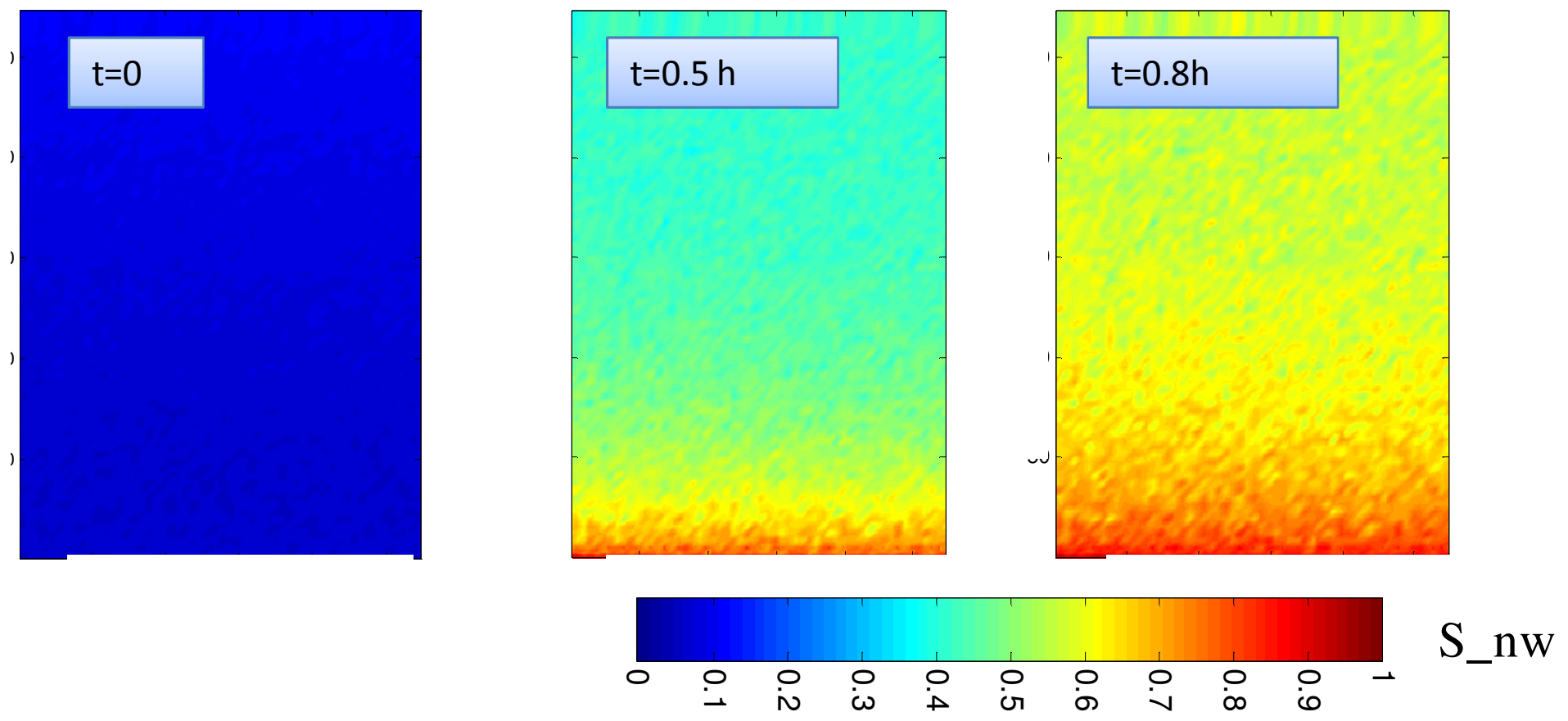

**Figure 4. Effective saturation of non-wetting phase evolution**

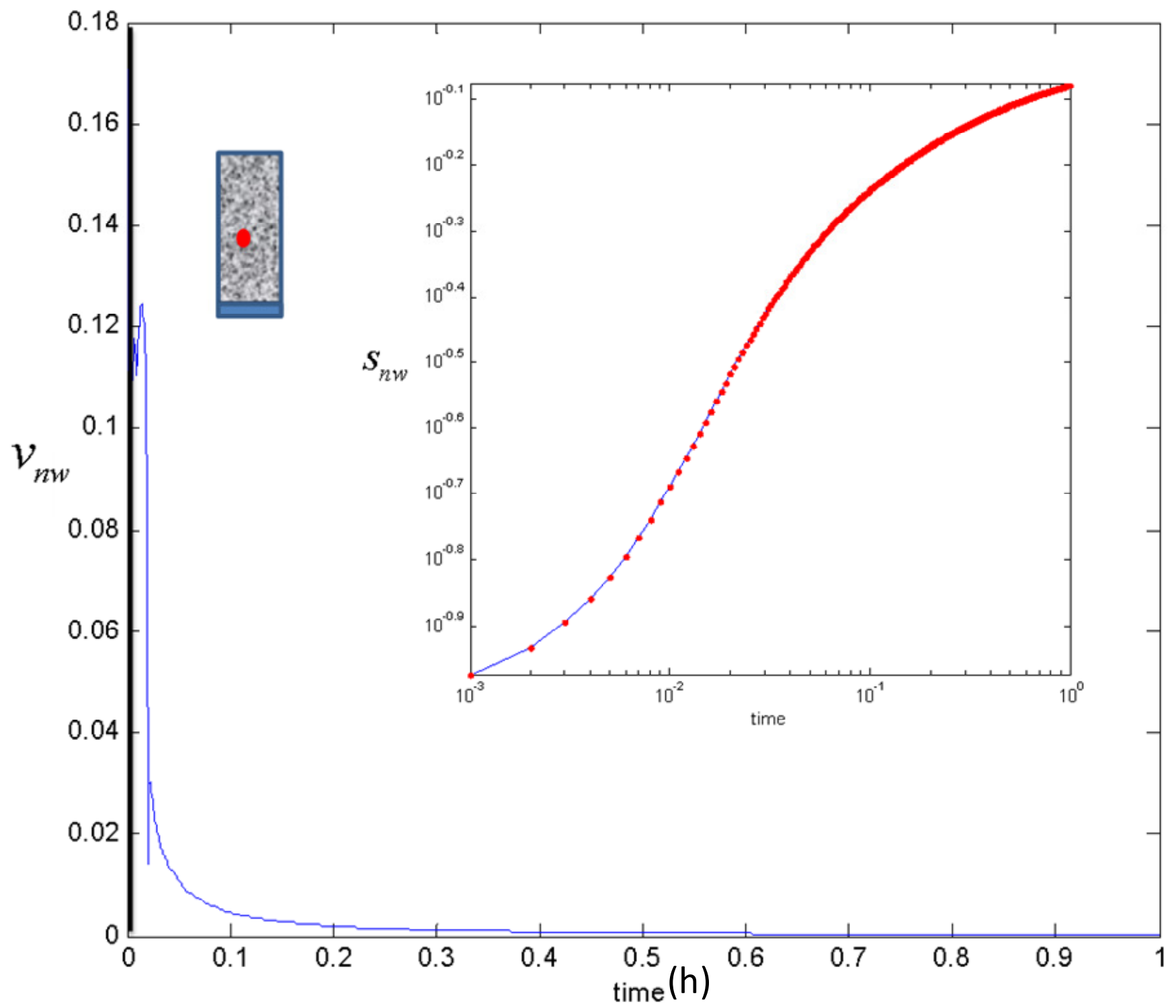

**Figure 5. Absolute air-velocity fluctuation (inset: effective saturation of non-wetting phase evolution along time)**

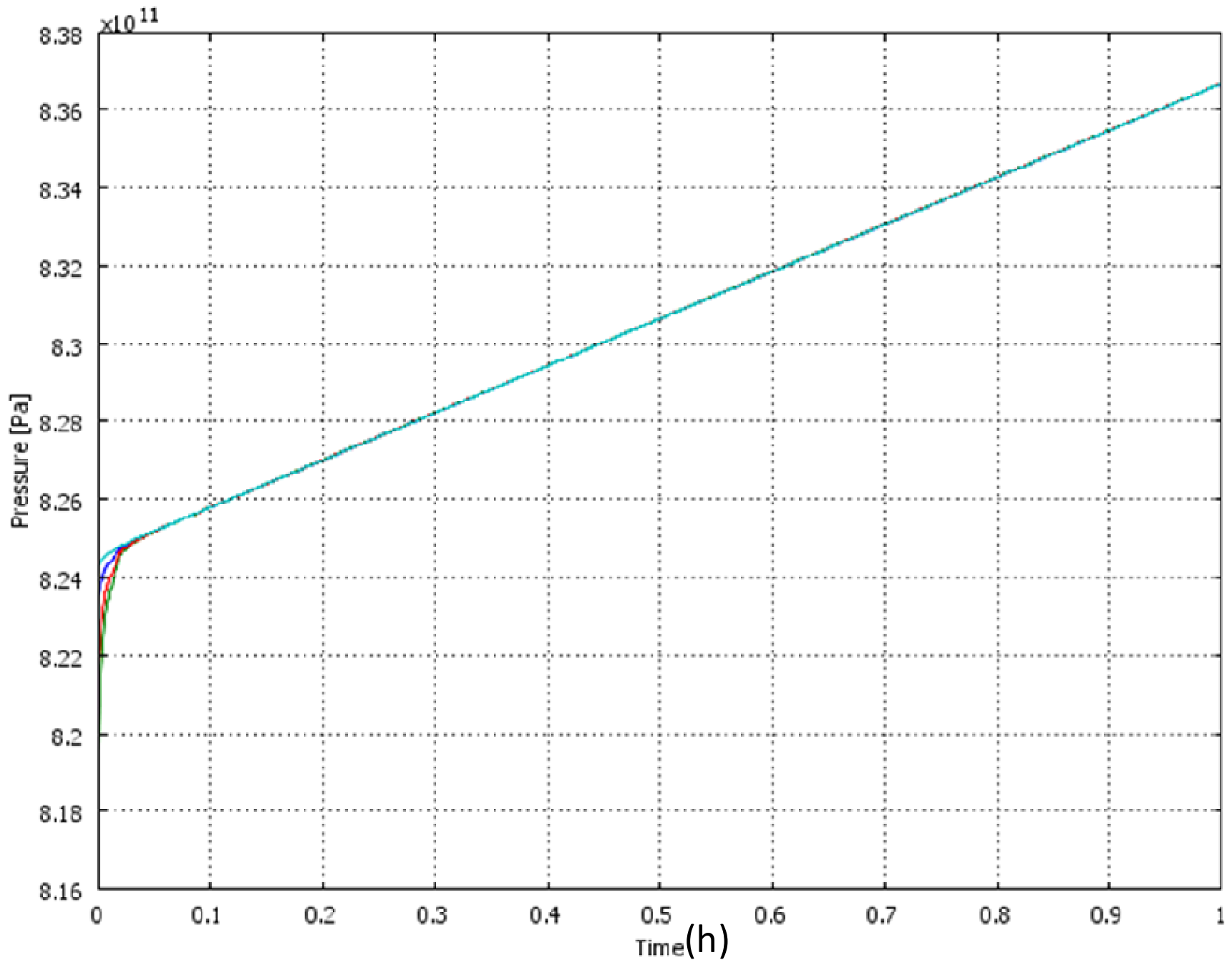

Figure 6. Air pressure changes vs. time for four randomly selected points

Figures 4-6 represent the result of simulation with 1857 triangular elements for saturation of non-wetting phase ,absolute value of air-velocity and pressure of non-wetting phase evolution, respectively. Based on the later mentioned figure it seems in association with initial boundary condition a phase transition before 0.1 h can be observed. The point of phase transition is known as a critical point to include some basic terminology. Usually two types of phase transitions are distinguished, first order and second order. First order designates phase transition where the macroscopic states change in a discontinuous way upon passage through the critical point, and second order indicates phase transitions where the states change in a continuous way [23]. When a system passes through a second order transition; it may be left at that point. At this point the system is like a pencil balanced on its end (only small variation may be observed). We cannot tell which direction it will fall, but a small perturbation can send it falling in a certain direction. In our case, the phase transition can be pronounced around 0.074 *h* where overall signals of the variables states show an indented behaviour change.

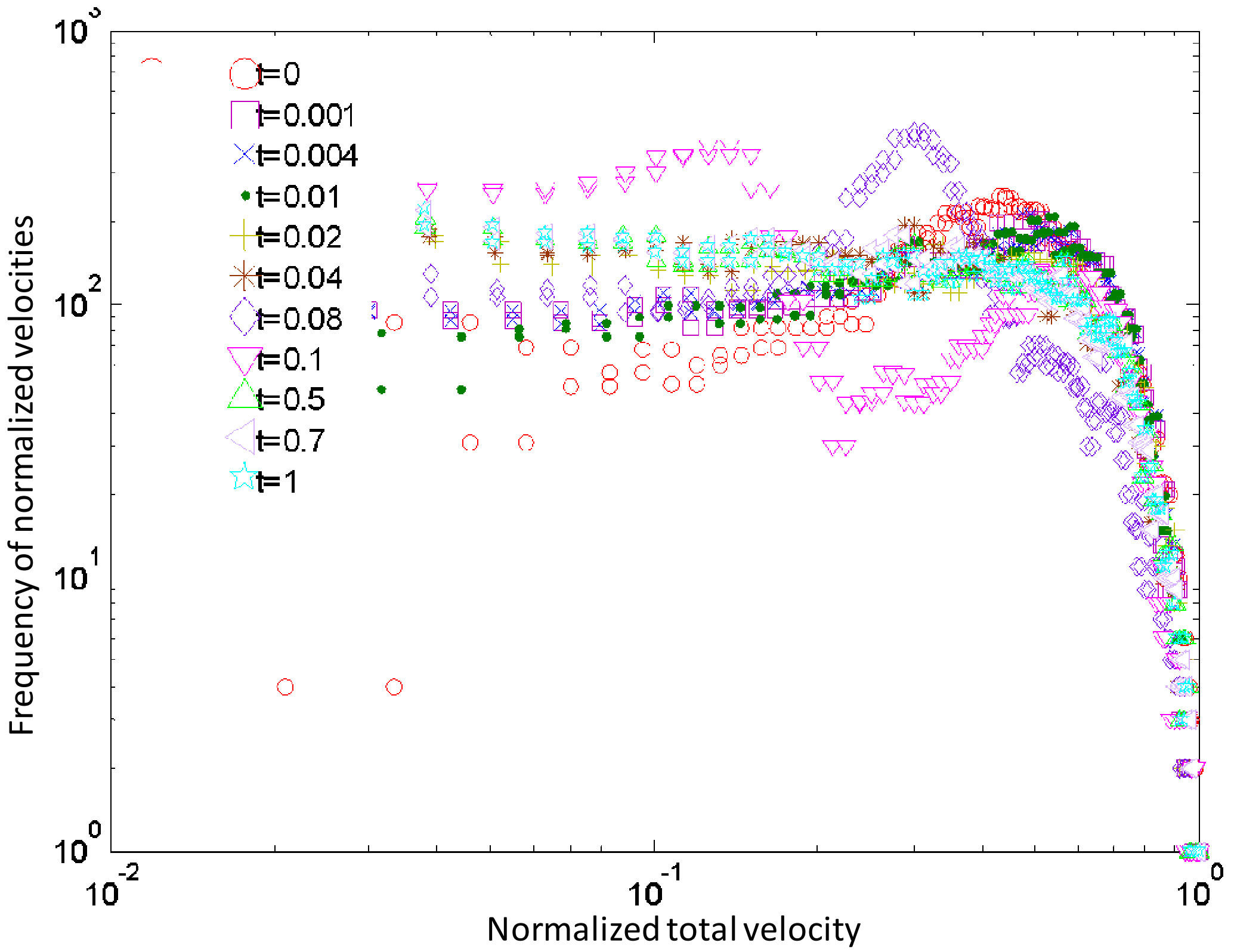

Figure 7. frequency distribution of the total velocity over time in log-log coordinate

Depicting of normalized (between 0 and 1) velocity frequency (Fig.7) shows the general overall probability patterns of locally scaled- velocity. Roughly speaking, It can be followed that the frequency of displacements rates of non-wetting phase is well approximated by a truncated power-law [23, 24]: $N(v) = (v + v_0)^{-\beta} \exp(-v/\kappa)$ ; where $v_0 = 0.05; \beta = 1.5; \kappa = 0.85$. This equation suggests that air particles motion follows approximately a truncated Levy flight (Levy processes), indicating that, despite the diversity of their travel history rate, the scaled rate of displacement of non-wetting phase particles follow reproducible patterns. The main property of the levy walk procedures is obeying from the infinite variance despite of the Gaussian process which has a finite variance. However, extension of Levy flight to truncated Levy flight transfers the infinite variance to finite variance [25-26].

## 3. Complex networks

As we mentioned in the previous section, the emerged patterns are following relatively complex forms. Such complex forms come out in the signals and time series formats. Recently, one of the methods for analysis of signals (information patterns) which has been considered is employing of complex networks over the time-space (spatial/temporal). Different methodologies have been proposed to capture the signal evolution based on the networks formation [31].

A network (graph) consists of nodes and edges connecting them. To set up a network, we consider each profile (in y-direction- saturation or pressure of non wetting phase) as a node. To make edge between two nodes, a relation should be defined. Several similarity or metric spaces has been proposed for a construction of a proper network. The main point in selection of each space is to explore the explicit or implicit hidden relation among different distributed elements of a system. In this study we will use p-value criterion (a matrix for testing the hypothesis of no correlation) related to correlation measurement and Euclidean distance over the non-wetting saturation profiles. For each pair of signals (profiles) $V_i$ and $V_j$, containing $L$ elements (pixels) the correlation coefficient can be written as [30-31]:

$$C_{ij} = \frac{\sum_{k=1}^{L}\left[V_i(k) - \prec V_i \succ\right].\left[V_j(k) - \prec V_j \succ\right]}{\sqrt{\sum_{k=1}^{L}\left[V_i(k) - \prec V_i \succ\right]^2}.\sqrt{\sum_{k=1}^{L}\left[V_i(k) - \prec V_i \succ\right]^2}} \quad (14)$$

where $\prec V_i \succ = \frac{\sum_{k=1}^{L} V_i(k)}{L}$. Obviously, $C_{ij}$ is restricted to the $-1 \le C_{ij} \le 1$, where $C_{ij} = 1, 0 \;\; and \;\; -1$ are related to perfect correlations, no correlations and perfect anti-correlations, respectively. Each p-value is the probability of getting a correlation as large as the observed value by random chance, when the true correlation is zero. The p-value is computed by transforming the correlation to create a t- statistic having N-2 degrees of freedom, where N shows number of profiles (samples) [27]. If $p$ is small-here less than $p \le \xi = 0.05$- then the correlation $C_{ij}$ is significant. Euclidean distance is given as:

$$d_{ij}=\sqrt{\sum_{k=1}^{L}(V_i(k)-V_j(k))^2} \tag{15}$$

Selection of threshold ($\xi$) is a challengeable discussion that can be seen from different view. Choosing of such constant value may be associated with the current accuracy at data accumulation where after a maximum threshold the system loses its dominant order. In fact, there is not any unique way in selection of constant value, however, preserving of general patterns of evolution must be considered while the hidden patterns (in our study come out from the solution of two-PDEs based on a two-phase flow constitutive model) can be related to the several characters of a networks . These characters can express different facets of the relations, connectivity, assortivity (hubness), centrality, grouping and other properties of nodes and/or edges. Generally, it seems obtaining stable patterns of evolution (not absolute) over a variation of $\xi$ can give a suitable and reasonable formed network. For the Euclidean distance we set $d_{ij} \geq \xi = 0.7 d_{ij}^{\max}$ . Consider with this definition we are investigating dissimilarity of profiles over the metric space while for former metric the variations of profiles related to each other are measured.

Let us introduce some properties of the networks: clustering coefficient ($C$) , the degree distribution ($P(k)$) and average path length ($L$ ). The clustering coefficient describes the degree to which $k$ neighbors of a particular node are connected to each other. Our mean about neighbors is the connected nodes to the particular node. To better understanding of this concept the question "are my friends also friends of each other?" can be used. In fact clustering coefficient shows the collaboration between the connected nodes to one. Assume the $i^{th}$ node to have $k_i$ neighboring nodes. There can exist at most $k_i(k_i-1)/2$ edges between the neighbors (local complete graph). Define $c_i$ as the ratio

$$c_i = \frac{actual \quad number \quad of \quad edges \quad between \quad the \quad neighbors \quad of \quad the \quad i^{th} \quad node}{k_i(k_i-1)/2} \tag{16}$$

Then, the clustering coefficient is given by the average of $c_i$ over all the nodes in the network [7,10]:

$$C = \frac{1}{N}\sum_{i=1}^{N} c_i \,. \qquad (17)$$

For $k_i \le 1$ we define $C \equiv 0$. The closer $C$ is to one the larger is the interconnectedness of the network. The connectivity distribution (or degree distribution), $P(k)$ is the probability of finding nodes with *k* edges in a network. In large networks, there will always be some fluctuations in the degree distribution. The large fluctuations from the average value ($< k >$) refers to the highly heterogeneous networks while homogeneous networks display low fluctuations [6]. The average (characteristic) path length *L* is the mean length of the shortest paths connecting any two nodes on the graph. The shortest path between a pair (*i, j*) of nodes in a network can be assumed as their geodesic distance $g_{ij}$, with a mean geodesic distance *L* given as below [11]:

$$L = \frac{2}{N(N-1)}\sum_{i<j} g_{ij}, \qquad (18)$$

where $g_{ij}$ is the geodesic distance (shortest distance) between node *i* and *j*, and *N* is the number of nodes. We will use a well known algorithm in finding the shortest paths presented by Dijkstra [28]. Based on the mentioned characteristics of networks two lower and upper boards of networks can be recognized: regular networks and random networks (or Erd˝os-Renyi networks [5-6]). Regular networks have a high clustering coefficient (C ≈ 3/4) and a long average path length. Random networks (construction based on random connection of nodes) have a low clustering coefficient and the shortest possible average path length. However Watts and Strogatz [30] introduced a new type of networks with high clustering coefficient and small (much smaller than the regular ones) average path length (this is called SW property).

To extraction of saturation networks we set up X=273*Y=273 points on the constrained area (only upper part-see figure 1). During the evolution of system, in 10 time-points using aforementioned method, complex networks along Y-direction and on the non-wetting phase were elicited. Consider that the numbers of evaluated points for construction of networks are smaller and bigger than the number of (finite) elements and heterogeneous elements which give the solution. This implies a limited correlation around each granule (node) with neighbourhood nodes. This can be followed by the evolution of adjacency matrix visualization where for correlation measurement and Euclidean distance have been depicted at figure 8. Appearance of a thin layer of connections (disconnections for Euclidean distance) around diagonal of matrix along successive time steps suggests a correlation around each node. It seems that the variation of the emerged network over the p-value measurement is insignificant whereas networks based on dissimilarity of Euclidean metric shows much higher variations. Evolution of number of edges, mean clustering coefficient and average path length can be followed in Figures 9 -11, respectively. As one can ensue the pattern and general trend of clustering coefficient and edges are similar for each case and average path length shows an opposite variation with the later properties. Then it can be interpreted, in mean view, increasing or decreasing of edges rate is near to forming or decaying of triangles (clustering coefficient). Also, high clustering coefficient and low average path length (especially for Euclidean measurement at last time steps) can be distinguished (property of small-world networks) [6, 27]. Notice the fluctuation of properties of saturation networks are not as well as the change of saturation (Fig.5), but is so similar to absolute value of air velocity. Roughly speaking, one can infer from comparison of velocity and $L, < c_i >, K$ that maximum velocity due to numerical solution coincides with $< c_i >^{t=0.005-0.01}_{correlation}, K^{t=0.005-0.01}_{correlation}$ . It is noteworthy before t<0.02 the properties of two distances are reveres each other but after this step the general trends are same so that we can write:

$$\begin{aligned}
&\max\{K,C,L\}^{\text{Euclidean}} = t\{0.08, 0.08, 0.02\} \\
&\min\{K,C,L\}^{\text{Euclidean}} = t\{0.02, 0.02, 0.08\} \\
&\max\{K,C,L\}^{\text{p-value}} = t\{0.01, 0.01, 0.02\} \\
&\min\{K,C,L\}^{\text{p-value}} = t\{0.02, 0.02, 0.01\}
\end{aligned}.$$

Consider that maximum (minimum) value of $K$ and $C$ are adapted to minimum (maximum) $L$ and vice versa. If we consider t=0.02 as transition point as has been depicted in figure 5; this gives the point (or stage) with lowest stability which may indicate lowest correlation (t=0.02). The temporal evolution of clustering coefficient and edges over Euclidean distance so approve this stage is a much more chaotic step where after this point the patterns of spatial and temporal characters of structure complexity approximately reach a quasi stationary step. To understand the path evolution of saturation profiles, we construct state diagram of $k_i - c_i$ for each case (Fig.14). For correlation distance (Fig.14 a) nodes generally (without considering time effects) lies from higher clustering coefficient to a congested area where in this area (can be called equilibrium place) we have $100 < k_i^{core} < 120; 0.4 < c_i^{core} < 0.45$ .

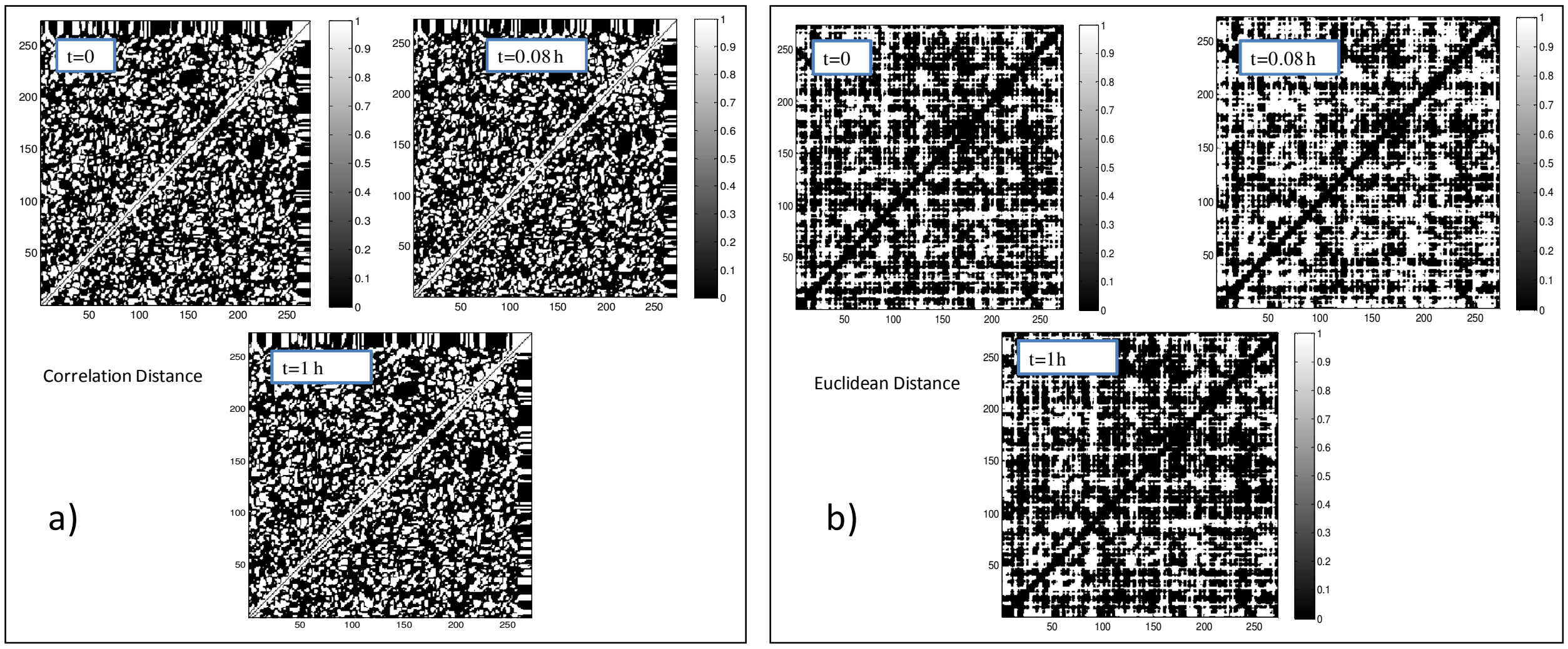

Figure 8. Evolution of adjacency matrix visualization of non-wetting phase saturation along time: based on a) p-value of Correlation measurements and b) Euclidean Distance

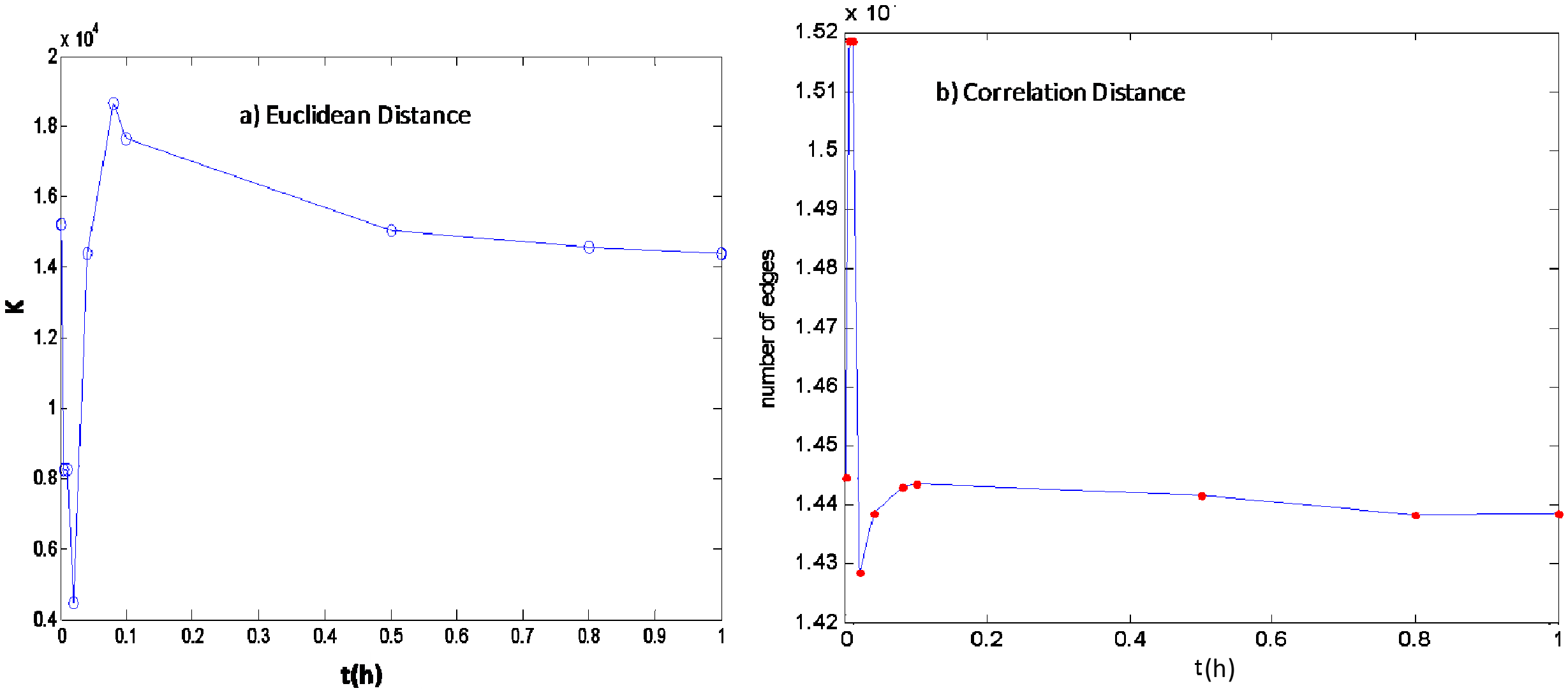

Figure 9. Variations of number of edges in complex saturation networks vs. time: based on a) Euclidean Distance and b) p-value of Correlation measurements

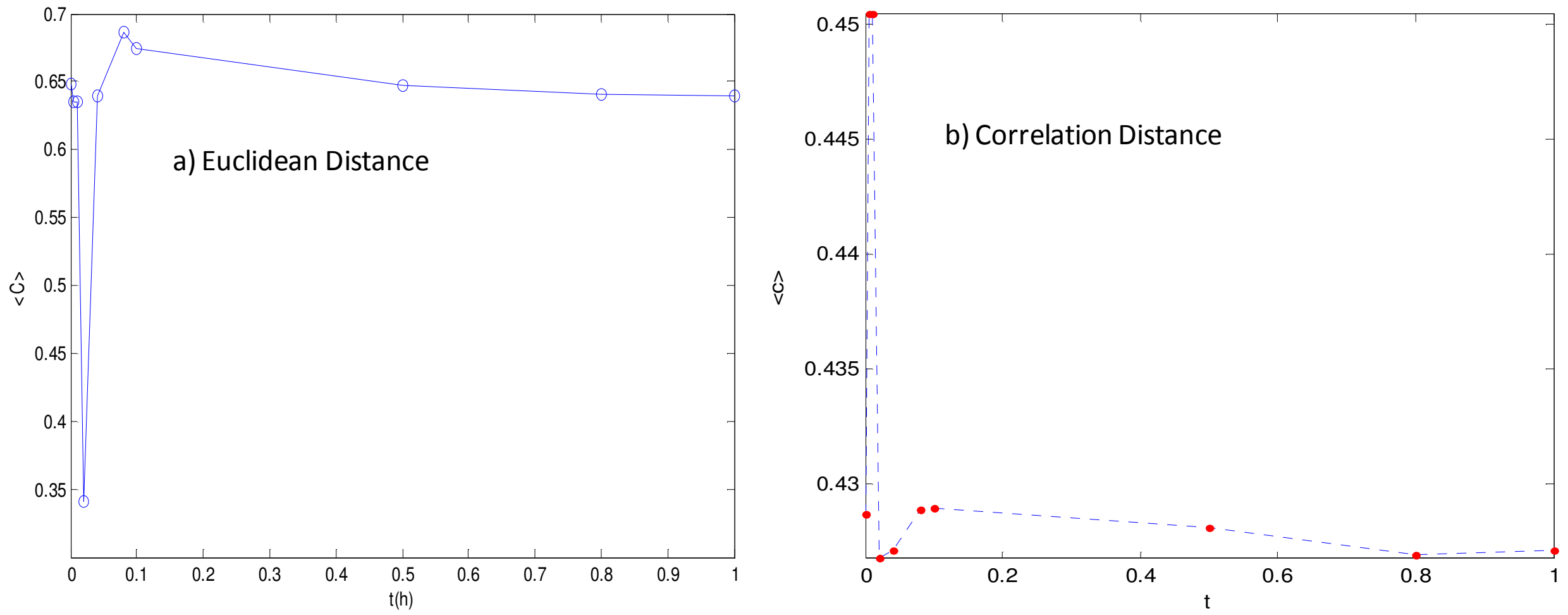

**Figure 10. Variations of mean clustering coefficient in complex saturation networks vs. time: based on a) Euclidean Distance and b) p-value of Correlation measurements**

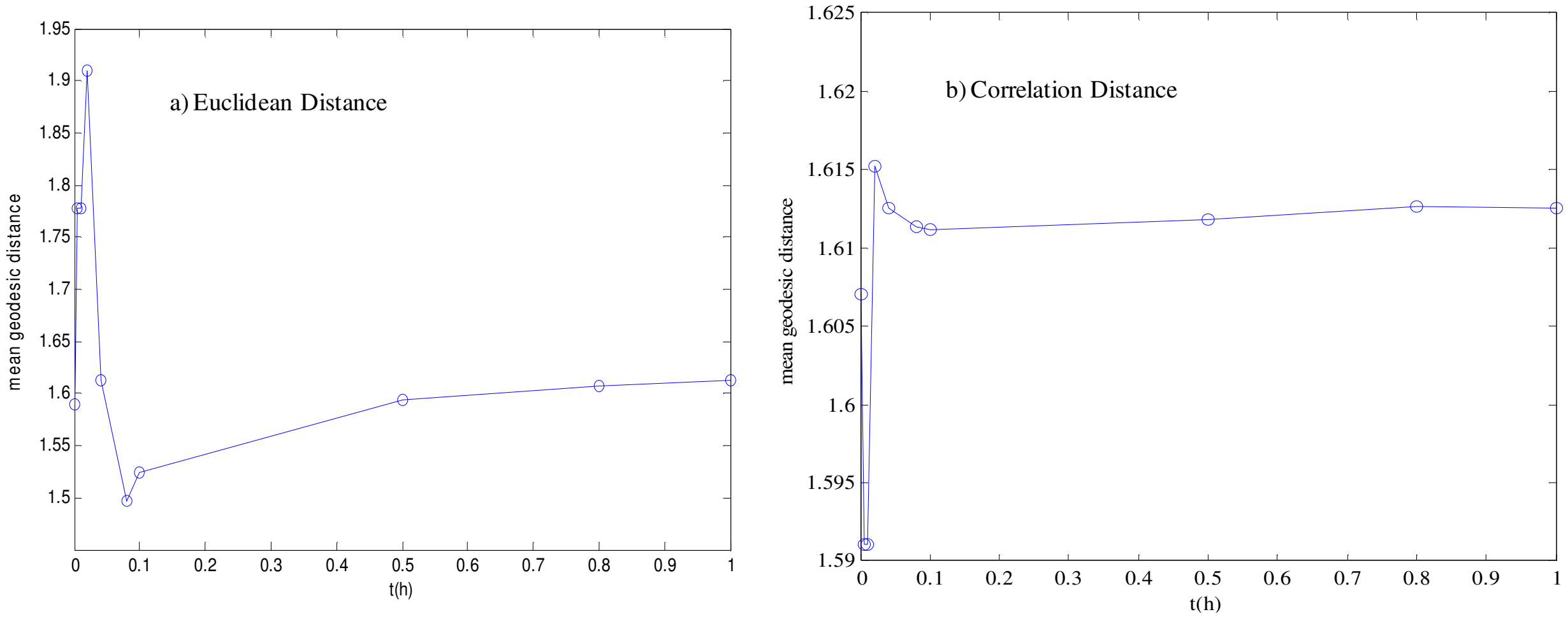

**Figure 11. Evolution of mean geodesic distance (mean length of the shortest paths) of complex saturation networks along time: a) Euclidean Distance and b) p-value of Correlation measurements**

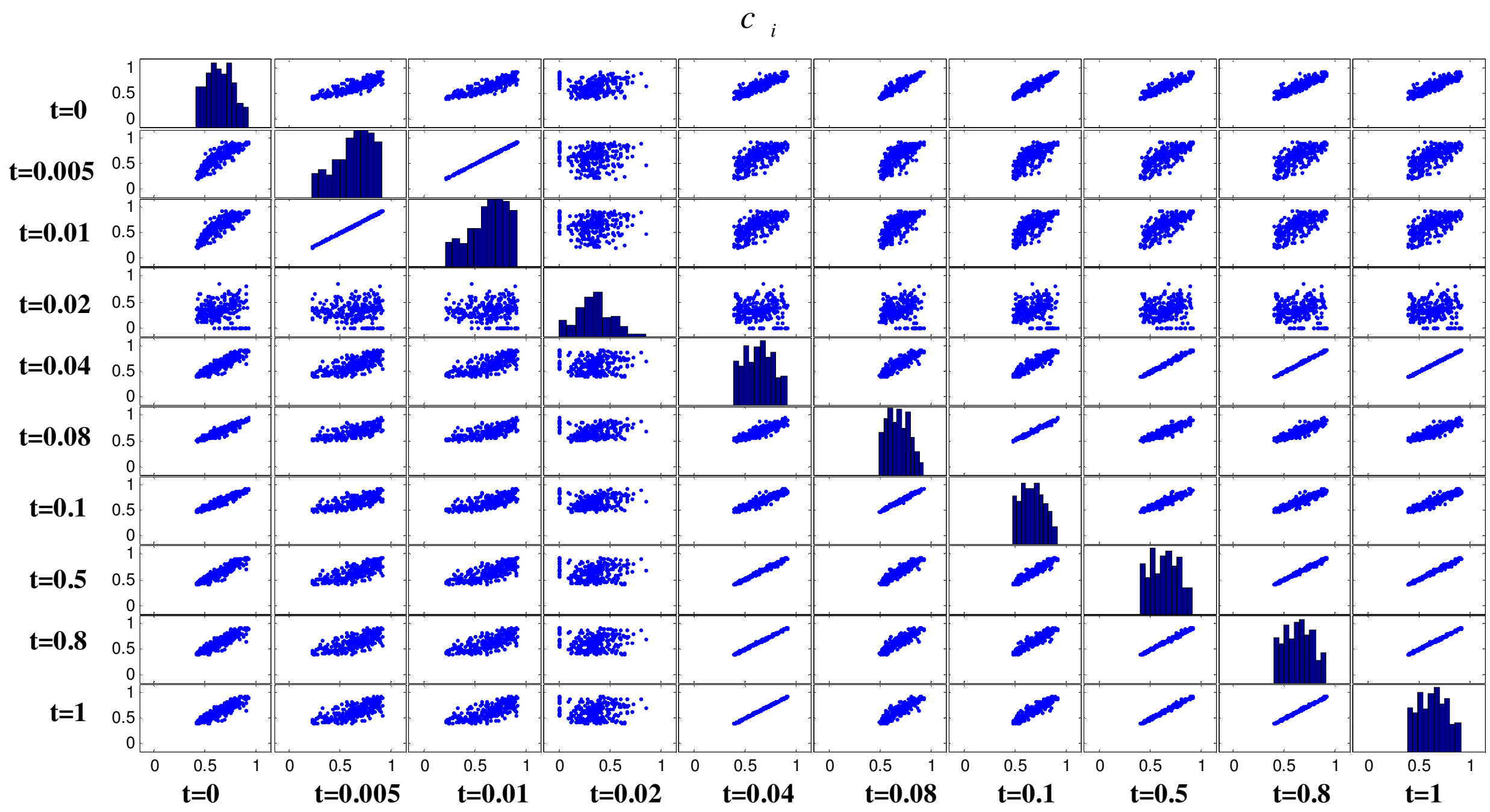

Figure 12. Temporal change of $c_i$ related to different reference point s in association with Euclidean Distance

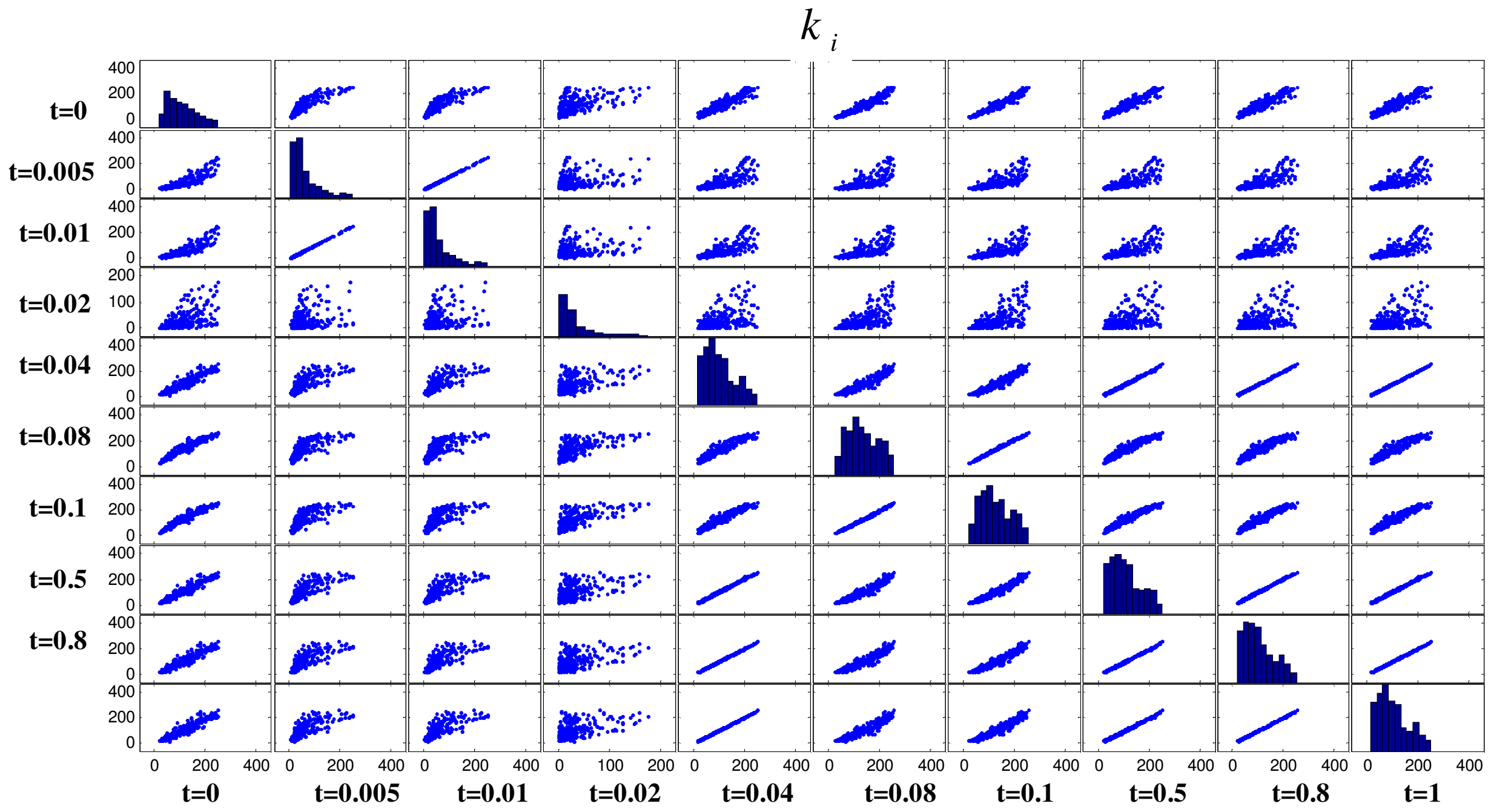

Figure 13. Temporal change of $k_i$ related to different reference points (time) in association with Euclidean Distance

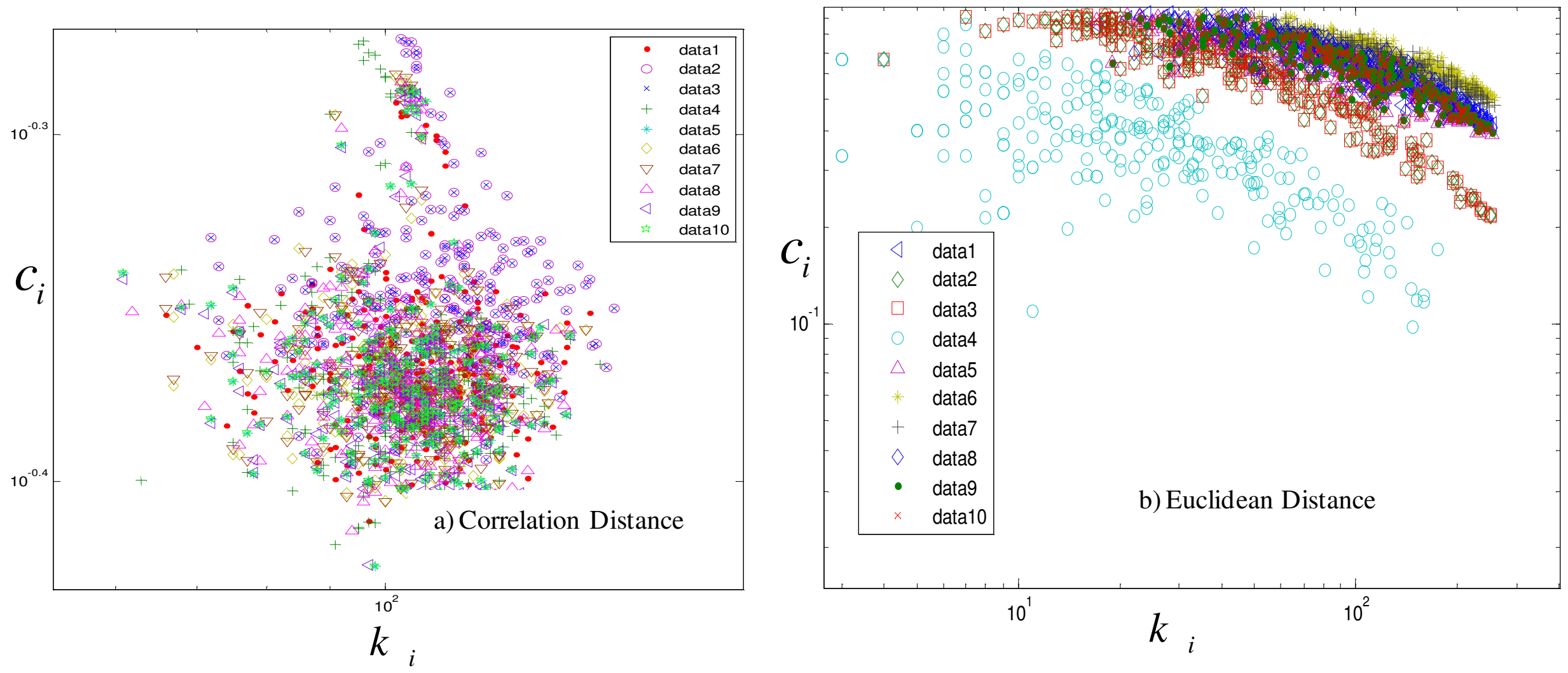

**Figure 14. Diagram of structure complexity space over two variables of networks $k_i$ and $c_i$ : a) p-value of Correlation, b) Euclidean Distance-data 1 to 10 are the sampling values within the time steps.**

Then, roughly speaking, the path evolution of saturation profiles in correlation measurement shows a narrow variation over the complexity structure space so that at stationary state the core of space can be imagined as a point of absorption. It must be considered that the discrimination of time steps over this evolutionary grape shape path is difficult. Despite of the former metric, Euclidean distance can give irregularities over time and space (Fig14b). Figure 14.b shows in t=0.02 there is distinguished chaotic separately different from overall way of change of patterns which is matched with pervious results. Also, the way of evolution can be estimated by: $c_i \approx 2.61 k_i^{-0.32}$. One may modify this relation with a truncated power law. In another view and with refer to our results drawn in Figure 14; we can conclude that the correlated and dissimilar patterns of saturation profiles (at a closed system) in a proper space of networks may be assumed as synchronizing oscillators. In other word, based on anharmonic oscillators' theory and the governing non linear differential equations we can write (for Figure 14):

$$\frac{d^2 c_i}{dt} = -\omega^2 \sin c_i(t) - \alpha \frac{dc_i}{dt} + f \cos \omega t,$$

in which $\omega$, $\alpha$ and $f$ are constant values.

Notice that with increasing of edges the communicability of system on this space is decreasing which shows the dissimilarity patterns tend to growth over all on nodes rather that the neighbourhoods of nodes. So, comparison with the distribution of edges indicates that nodes with high correlation and dissimilarities are decaying as well as triangles densities around each node (Figs.15&16). This implies that low dissimilar patterns (small $k_i$) have higher communicability. Let us investigate the temporal structure complexity. Two immediate solution can be presented in first situation matrix of connections along evolution of time is extracted. In this case spatial connection of nodes is replaced with temporal evolution of same node with same node at next (or pervious) time steps or other nodes. Consider a reference time step ($t^{ref.}$) then we can make the following matrix among $t^{ref.}, t^{ref.} + \Delta t$:

$$M^{t^{ref.}, t^{ref.}+\Delta t} = t^{ref.}\left\{ \overbrace{\begin{bmatrix} . & . & . \\ . & m_i & . \\ . & . & . \end{bmatrix}}^{t^{ref.}+\Delta t}_{N\times N} \right. \quad ; m_i \in \{0,1\}; i = 1,..,N \tag{19}$$

where $m_i$ is the connection state of element $i$ (position $i$) due to one relation along a time interval. So, a multiple array over different time references can cover completely both spatial ($\Delta t = 0$) and temporal evolution of structure complexity such:

$$B = \begin{bmatrix} M^{(t_0,t_0+\Delta t)} & . \; . & M^{(t_f,t_f+\Delta t)} \\ . & . \; . & . \\ . & . \; . & . \\ M^{(t_0,t_0+n\times\Delta t)} & . \; . & M^{(t_f,t_f+n\times\Delta t)} \end{bmatrix}_{n\times n} \quad ; n = 1 + {}^{t_f - t_0}\!\!\diagup_{\Delta t} \tag{20}$$

in which $n$ is the time granulation. Consider for $\Delta t = 0$, we can see only evolution of spatial complexity on the assumed time and in collective way as is usual in networks terminology: $[B]_{1\times n}$. The mentioned method can give general complexity patterns of time - space and interwoven structures, explicitly. Analysis of the attributes and properties of the $B$ is out of the scope of this study. Second case is an implicit (more qualitatively) view to time-based complexity. In this state, the spatial characteristics ($k_i, c_i$) of the system at several time steps will be compared (see figures 12 and 13).

In another procedure we construct air-velocity networks. The procedure is as well as correlation distance, only the number of nodes has decreased to 93. Three distinguished behaviours due to the nodes and edges variation are recognized, approximately (Fig.17 a, Fig.18a): 1) decreasing of nodes where the numbers of edges are approximately fixed (t= [0.01, 0.02]) compatibility in variations and 3) vertices are constant while sensitive variation in links can be perused (t=[0.3,1]). The behaviour of emerged network especially after (t>0.2) can be estimated by the properties of small-world networks. The shape of the degree distribution is similar to that of a random graph. It has a pronounced peak at $\prec k \succ_{t>0.2} \approx 8$ while the topology of the network is relatively homogeneous and all nodes having approximately the same number of edges (Fig.18 b).

Another point can be inferred from the plotting of the reverse of mean clustering coefficient so that in semi-logarithmic coordinate (Fig17.c) it follows a phase-transition (first order) function that can be followed by a sigmoid function [22]:

$$\prec 1/c \succ \propto (1+e^{-\beta^{(1+\frac{Lnt}{1+\delta Lnt})}}) \tag{21}$$

where $\beta$and $\delta$are the regulator parameters which determine the declining rate. Then, the reverse of mean clustering coefficient for velocity networks gives an acceptable phase change criteria.

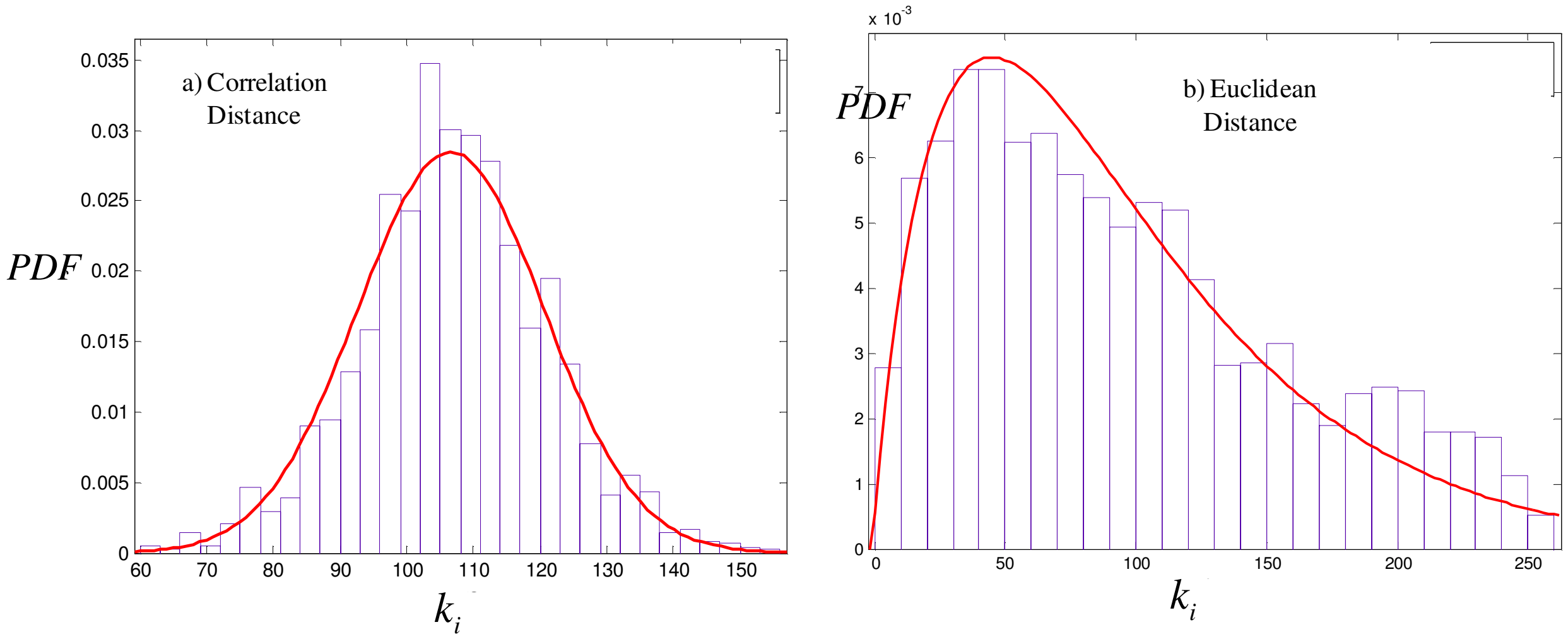

Figure15. Edge distribution of complex saturation networks: based on a) p-value of Correlation measurements and b) Euclidean Distance

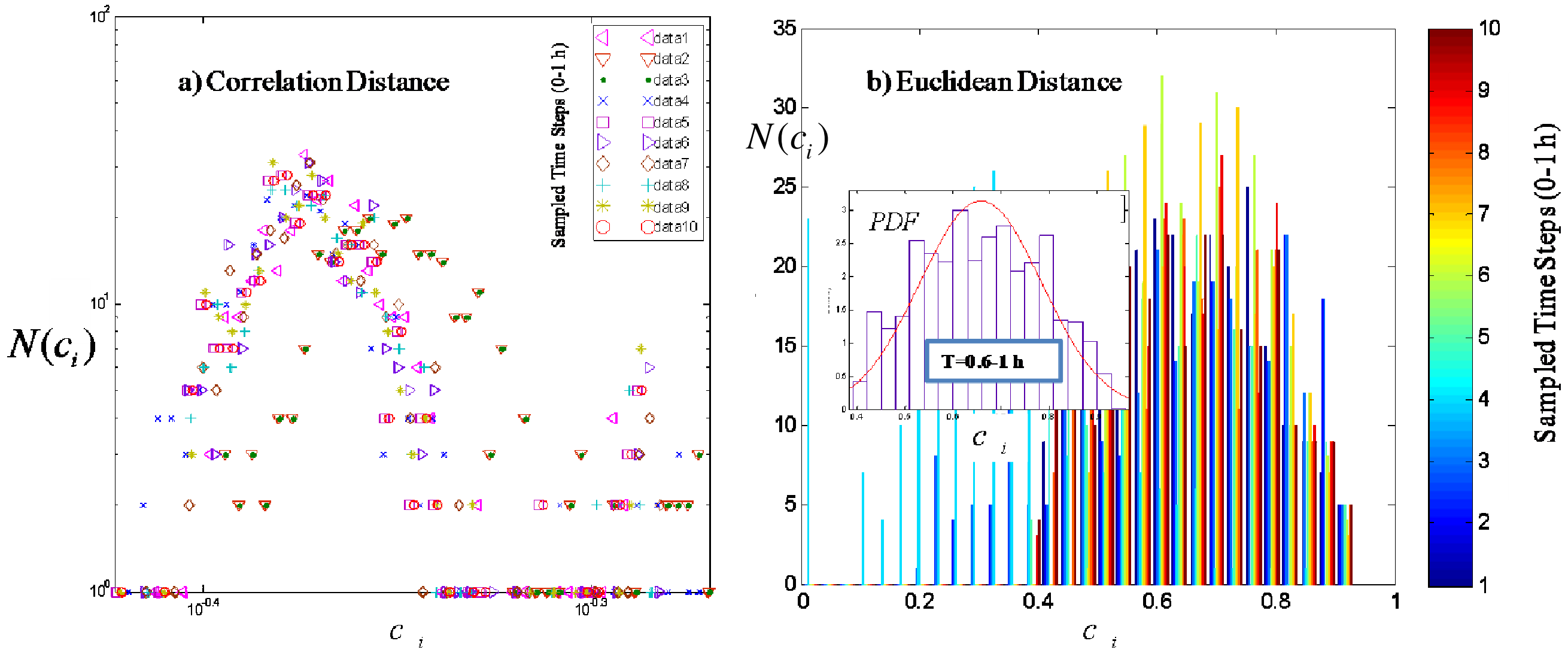

Figure 16. Evolution of mean clustering coefficient distribution along time: based on a) p-value of Correlation measurements and b) Euclidean Distance (inset: probability density function at 0.6<t <1 h)

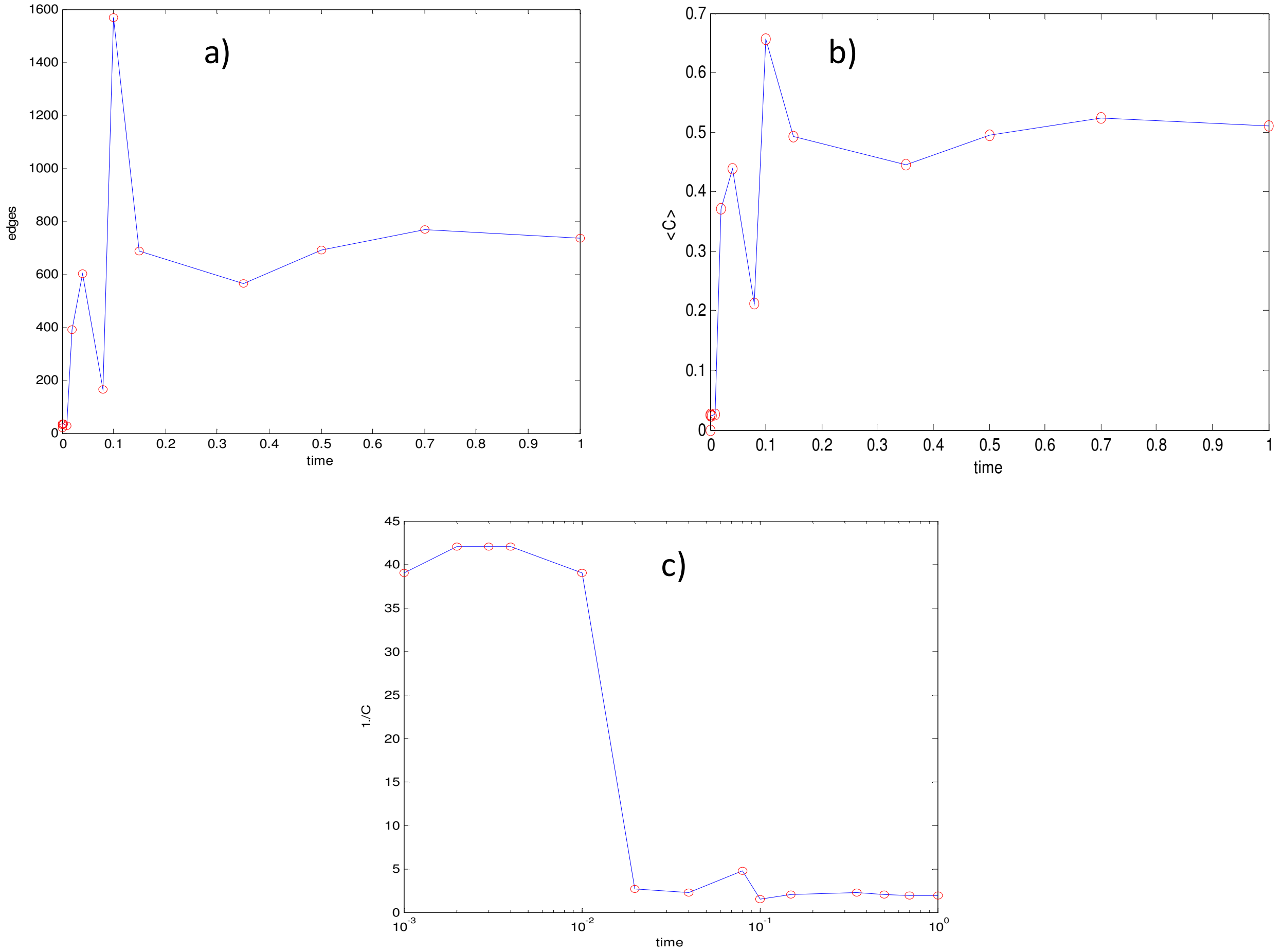

Figure 17. a) Evolution of total number of edges for air velocity networks , b) Mean cluster coefficient variation and c) inverse of Mean cluster coefficient variation

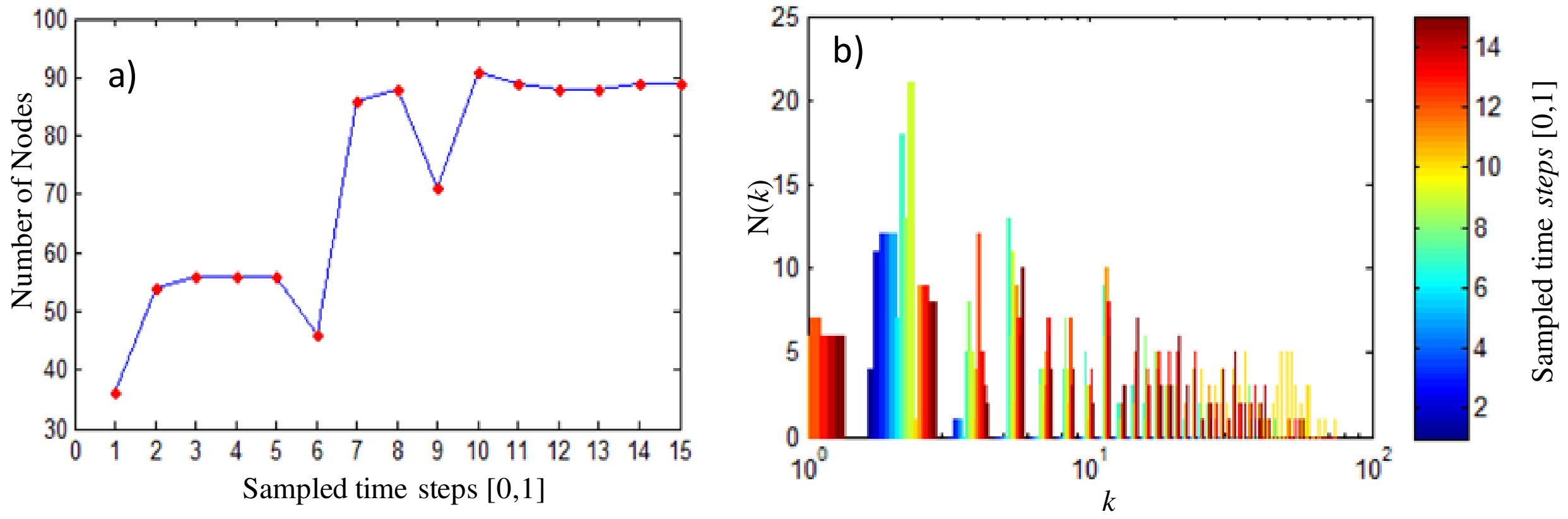

Figure 18.a) Number of nodes-time ;b) frequency of edges evolution along time for velocity (of air) networks

## 4. Conclusion

Analysis of two phase flow especially in heterogeneous media, has allocated a noticeable research literature in porous media area. To take in to account the complexity due to heterogeneity in different parameters which are characterizing the general reaction (s) of the field (s), different methodologies have been proposed. The aim of this study was to investigate the appeared complexity of two-phase flow (air/water) in a heterogeneous soil. The supposed filed was under the small range of air pressure variation. By considering the capillary pressure-saturation, permeability functions, as constitutive relations and governing equations (obtained separately for evolution of the model's wetting and non-wetting phase parameters) the unknown parameters can be estimated. In this way, using finite element method the role of multi-heterogeneity the successive change of different variables (such relative permeability, saturation, capillary pressure etc) was analysed. Based on the different similarity measurements (i.e., correlation, Euclidean metrics) over the properties of the emerged patterns of the complex networks were recognized. In this way, the path of evolution of the supposed system was illustrated based on the state space of networks either in correlation and Euclidean measurements. The results of analysis showed in a closed system the designed complex networks approach to small world network property where the mean path length and clustering coefficient are low and high. As another result, we tried to scale the evolution of macro -states of system (such mean velocity of air or pressure) with characteristics of structure complexity of saturation.